\documentclass{amsart}

\usepackage{amsmath}
\usepackage{amssymb}
\usepackage{verbatim}
\usepackage{cite}

\DeclareMathOperator*{\argmax}{arg\,max}

\usepackage[a4paper, total={6in, 8in}]{geometry}
\title{Probabilistic Approach for Evaluating Metabolite Sample Integrity\\}
\author{Barry M. Slaff, Shane T. Jensen, and Aalim M. Weljie}

\begin{document}

\begin{abstract}

The success of metabolomics studies depends upon the ``fitness'' of each biological sample used for analysis: it is critical that metabolite levels reported for a biological sample represent an accurate snapshot of the studied organism's metabolite profile at time of sample collection. Numerous factors may compromise metabolite sample fitness, including chemical and biological factors which intervene during sample collection, handling, storage, and preparation for analysis. We propose a probabilistic model for the quantitative assessment of metabolite sample fitness. Collection and processing of nuclear magnetic resonance (NMR) and ultra-performance liquid chromatography (UPLC-MS) metabolomics data is discussed. Feature selection methods utilized for multivariate data analysis are briefly reviewed, including feature clustering and computation of latent vectors using spectral methods. We propose that the time-course of metabolite changes in samples stored at different temperatures may be utilized to identify changing-metabolite-to-stable-metabolite ratios as markers of sample fitness. Tolerance intervals may be computed to characterize these ratios among fresh samples. In order to discover additional structure in the data relevant to sample fitness, we propose using data labeled according to these ratios to train a Dirichlet process mixture model (DPMM) for assessing sample fitness. DPMMs are highly intuitive since they model the metabolite levels in a sample as arising from a combination of processes including, e.g., normal biological processes and degradation- or contamination-inducing processes.  The outputs of a DPMM are probabilities that a sample is associated with a given process, and these probabilities may be incorporated into a final classifier for sample fitness.

\end{abstract}

\maketitle

\section{Introduction} \label{introduction}

Quantitative analysis of metabolite levels in biofluids and tissues has become a fruitful approach in medical and translational research. The success of such studies depends upon the ``fitness'' of each biological sample used for analysis. Specifically, it is critical that metabolite levels reported for a biological sample represent an accurate snapshot of the studied organism's metabolite profile at time of sample collection. Between sample collection and final analysis, numerous factors may compromise metabolite sample fitness, including chemical (e.g. thermodynamic) and biological (e.g. bacterial) factors which intervene during sample collection, handling, storage, and preparation for analysis. We propose a probabilistic model for the quantitative assessment of metabolite sample fitness. The proposed model may be implemented as a computational tool which uses the measured metabolite profile from a biological sample to estimate the sample's fitness for inclusion in metabolomics analyses.\\

In the present work, we present an approach for developing a probabilistic model of metabolite sample fitness. In Section \ref{inputs} we discuss the proposed analytical methods for collecting metabolite sample data. The proposed analytical platforms are nuclear magnetic resonance (NMR) spectroscopy with a targeted profiling approach for quantitation together with ultra-performance liquid chromatography coupled to mass spectrometry (UPLC-MS). These are established methods for the acquisition of quantitative metabolite data from biofluid and tissue samples. Additionally, we discuss the critical data processing steps of normalization, centering, and scaling. Finally, we discuss approaches to feature selection for the final model-building process.\\

In Section \ref{principle} we discuss the principle of metabolite sample fitness. Critical to the modeling process is a quantitative definition of sample fitness: how should fitness be assessed? Existing studies suggest the possibility of identifying metabolites that change significantly in concentration (``changing metabolites'') and metabolites which remain relatively stable (``stable metabolites'') during sample storage at different temperatures. We propose that at least one changing-metabolite-to-stable-metabolite ratio might be identified for each biological matrix, and tolerance intervals will be computed to characterize the range of ratios among fresh samples. Therefore, a sample can be deemed fit or unfit based on whether its changing-metabolite-to-stable-metabolite ratios fall within the computed tolerance intervals for fresh samples. While this method offers a straightforward computation, it does not capture all the structure available in the data for differentiating fresh and degraded samples. Therefore, we propose that data classified using tolerance intervals may be used to train a probabilistic model capable of capturing additional structure in the data.\\

In Section \ref{models}, we propose approaches for modeling metabolite sample fitness based on the principle discussed in Section \ref{principle}. Key to our approach is development of a model which avoids incorrect parametric assumptions. Our main modeling approach is development of a Dirichlet process mixture model (DPMM) for each biological matrix for which we wish to assess sample fitness. DPMMs are highly intuitive since they model the metabolite levels in a sample as arising from a combination of processes including, e.g., normal biological processes and degradation- or contamination-inducing processes.  The outputs of a DPMM are probabilities that a sample is associated with a given process, and these probabilities will be incorporated into a final classifier for sample fitness. We also consider the use of conceptually simple non-parametric methods such as k-Nearest-Neighbor and kernel regression.

\section{Model Inputs: Data Acquisition and Processing} \label{inputs}

Nuclear Magnetic Resonance (NMR) Spectroscopy and Ultra-Performance Liquid Chromatography - Mass Spectrometry (UPLC-MS) are state-of-the-art analytical platforms of the acquisition of quantitative metabolite data from urine, plasma, serum, tissue, and other biological  matrices\cite{Keun2011, Beckonert2007, Zhou2012, Want2013, Benton2012}. Quantitative data obtained using NMR and UPLC-MS will be utilized to model and assess metabolite sample integrity.\\

\subsection{Data Acquisition} \label{acquisition}

Experimental design and data acquisition in metabolomics must avoid contamination of the data with systemic errors and variances that can compromise analyses\cite{Razali2011}. Targeted profiling\cite{Weljie2006, Athersuch2013} with NMR spectroscopy produces quantitative metabolite concentrations which are reproducible within\cite{Weljie2008, Tredwell2011} and between\cite{Keun2002, Ravanbakhsh2014} labs, with more variation when NMR probes and experimental parameters are not consistent\cite{Lacy2014}. A Design of Experiments (DoE) approach together with UPLC-MS yields reproducible metabolite data in quantitative and non-quantitative approaches\cite{Zheng2013, Xiao2012, Oberacher2009}. Appropriate measures will be taken in the experimental design to avoid biases arising from test subject selection and temporal factors (e.g. time of day of sample collection). Sample preparation and storage procedures will be tightly controlled apart from those varied deliberately as part of the experimental procedure for inducing sample degradation. Our investigation of metabolite sample integrity will inform existing domain knowledge regarding proper sample preparation and storage for metabolomics studies\cite{Dona2014, Emwas2014, Kamlage2014, Yin2013, Korman2012, Bruce2009, Saude2007} and biobanks\cite{Elliot2007, Dunn2008, Hebels2013}.\\

\subsection{Data Processing} \label{processing}

Normalization, centering, and scaling are essential steps for metabolomics data processing. Normalization is a critical step for purposes of comparing spectra acquired from samples with different levels of dilution\cite{Giraudeau2014, Ejigu2013, Wang2013}. This concern is particularly critical in the case of urine samples\cite{Slaff2014, Wang2013}. The simplest normalization method is total integral normalization, which assumes that all spectral peaks scale with sample dilution. We also consider probability quotient normalization\cite{Dieterle2006} (PQN), also called median-fold change normalization\cite{Veselkov2011} (MFC), which scales the peaks in each spectrum by one factor per spectrum so that the median fold-change between the peaks in each spectrum and corresponding peaks in a reference spectrum is 1. In contrast to total spectral normalization, PQN/MFC assumes that most rather than all spectral peaks scale with sample dilution. We also consider an additional normalization step which would minimize the distance from the sample vector to a modeled probability distribution. The metric used to evaluate nearness might be Euclidean distance as a default, or for example Mahalanobis distance\cite{Maesschalck2000} if the clusters are modeled as multivariate Gaussians. In the case of Gaussian clusters with diagonal covariance, this computation is analytically simple\cite{Geitz2007} and it is analytically or numerically computable in other cases. Total integral normalization (TIN) and PQN/MFC are widely used and have been studied comparatively in the context of both NMR\cite{Giraudeau2014} and LC-MS\cite{Ejigu2013} metabolomics data. It has been shown that use of TIN or PQN/MFC improves the results of comparing spectra relative to no normalization, while the optimal method varies between contexts\cite{Giraudeau2014, Ejigu2013, Wang2013}.\\

Prior to model-building with training data and classification with new data, the data may be mean-centered and scaled feature-by-feature. Scaling assumes that the features with the most variance are not necessarily the features with the most predictive value, since relatively abundant features tend to have greater variance. Several scaling methods widely used in metabolomics studies\cite{vandenBerg2006, Gromsky2014} include auto-scaling (each feature in the training data is scaled to have variance 1), Pareto scaling (each feature scaled so that its variance is the square-root of its initial variance), Variable stability (VAST) scaling\cite{Keun2003} (features with smaller coefficient of variation are given more weight), and range scaling (each feature is scaled by its full range). The optimal scaling approach has been found to be highly context-dependent, and the results of modeling depend significantly upon the scaling method utilized\cite{vandenBerg2006, Gromsky2014}.

\subsection{Feature Selection} \label{features}

It is often advantageous in data analysis contexts to utilize a subset of the acquired features or generate new features for the final modeling task. For example, eliminating irrelevant or noisy features can improve the predictive performance of any final model. Additionally, generating new features which are highly relevant to the prediction can improve model performance.\\

With respect to choosing subsets of acquired features for modeling, we consider the following:\\

\begin{enumerate}

  \item Since the consistently-detectable metabolites are known to differ across matrices for NMR\cite{Keun2011} and LC-MS\cite{Zhou2012}, features should be selected on a per-matrix basis, i.e. one feature set for human urine, one feature set for human serum, etc. The full panel of reliably-detectable metabolites will be profiled initially.\\
  
  \item A Design-of-Experiment approach\cite{Riter2005, Eliasson2012, Zheng2013} (DoE) may be used to experimentally narrow the matrix-specific feature lists. For example, metabolites common to the same molecular pathways may exhibit high co-linearity, which would be detectable with a DoE approach.\\
  
  \item In the case of a parametrized clustering approach, it may be useful to include only features which satisfy certain parametric constraints. For example, in a model which constructs multivariate Gaussian clusters, it may be interesting the maximize the Gaussian character of the joint distributions through feature selection. Multivariate Gaussian character can be assessed using an R package such as MVN\cite{KorkmazMVN}. This may be achieved by selecting only features (possibly after transformation) with univariate Gaussian character, which can be assessed with a univariate test for normality such as  Lilliefors\cite{LillieforsH, Razali2011, Moseley2013} with an R package such as nortest\cite{Gross2015}.\\
  
\end{enumerate}
  
It may be desirable to combine the initial metabolite features into new features for purposes of modeling metabolite sample fitness. Numerous feature selection methods have been utilized in omics studies, in particular in domains for which studies typically include many more features than samples\cite{Saeys2007, delosCampos2013}. Possible feature-combination approaches for modeling metabolite sample fitness include:\\

\begin{enumerate}

	\item Components from unsupervised spectral methods: utilizing principal component analysis (PCA) vectors (equivalently, singular vectors and values) or independent component analysis (ICA) vectors as features\cite{Bartel2013}.\\
	
	\item Latent vectors: utilizing latent vectors from partial least squares regression\cite{Wold2001} (PLS), canonical correlation analysis\cite{Hardoon2004} (CCA), or related spectral methods such as orthogonal PLS\cite{Trygg2002opls} (O-PLS) and O-PLS with discriminant analysis\cite{Bylesjo2007} for classification (O-PLS-DA). The orthogonalized methods remove systemic variation in the data unrelated to the response variables to improve interpretability. O-2-PLS\cite{Trygg2002o2} additionally yields two-way information about covariation and predictiveness between the observed and response variables. For these methods, the response variables could be, for example indicators of sample degradation such as time of sample exposure to non-freezing storage temperature. These methods have been used widely for metabolomics data analysis\cite{Korman2012, Madsen2010, Varmuza2009}.\\
	
	\item Feature clustering: The method of shrunken centroids\cite{Tibshirani2002}, which originated as a feature selection method in genomics, has been applied in a metabolomics context\cite{Sha2010, Chen2014} for choosing a subset of highly representative features. Other clustering approaches involve using mutual information\cite{Krier2007} or using a graph-theoretic approach\cite{Magendiran2014} to identify clusters and choose representative metabolite features.\\
	
	\item Multiple-testing framework for discovering significant features: kernelized support vector machines\cite{Srivastava2005}, k-nearest-neighbor\cite{Bhatia2010}, and classificaion trees\cite{Boulesteix2012} have been used together in a multiple testing framework to identify individual metabolite features\cite{Kim2008}. This approach is not widely used in the metabolomics literature but is attractive particularly since the false discovery rate can be controlled via the multiple testing framework.

\end{enumerate}

\section{Modeling Sample Fitness} \label{modeling}

We wish to distingish fit samples from unfit samples. In principle, a sample is fit for analysis if its measurable metabolite levels at time of analysis are very similar to its measurable metabolite levels at time of collection. According to this principle, a sample is fit if at time of measurement, it accurately captures a collection-time snapshot of the studied organism's metabolite profile. The original metabolite levels change over time due to intervention from chemical and biological factors during sample handling, storage, and preparation for analysis. Therefore, our central problems are the following:\\

\begin{enumerate}

	\item Quantify the degree of change in metabolite levels after collection for which the sample should no longer be considered ``fit'' for analysis. We follow the recommendation of Fraser et al\cite{Fraser1997}, now widely adopted for quality control reporting in clinical medicine\cite{Plebani2015}, and define three levels of sample fitness: optimal, desirable, and minimal.\\
	
	\item Construct a probabilistic model of metabolite sample fitness so that a sample can be accurately categorized as fit (optimal, desirable, or minimal) or unfit based upon its reported metabolite levels. The model will be trained on data for which we have followed the time-course of metabolite level changes from the absolutely fresh state to various states of degradation. The result of (1) will be used to label each training data point as fit (optimal, desirable, or minimal) or unfit. The trained probabilistic model will be used to predict the fitness or non-fitness of samples for which we have only one measurement of metabolite levels.\\
	
\end{enumerate}

Problem (1) is the subject of Section \ref{principle} and (2) is the subject of Section \ref{models}.

\subsection{Principle of Metabolite Sample Fitness} \label{principle}

Recent studies identify urine, blood, and plasma metabolites that change concentration significantly over hours and days in response to storage at above-freezing temperatures\cite{Emwas2014, Saude2007, Yin2013, Kamlage2014}. This degradation process transforms a sample from a state of fitness to unfitness. We hypothesize that ratios between changing metabolites and stable metabolites during storage can identify fresh vs. degraded samples. For this purpose we propose the use of tolerance intervals\cite{Wald1946, Brown2009} for characterizing changing-to-stable metabolite ratios in optimally-fit, desirably-fit, and minimally-fit samples. For each biological matrix, at least one changing/stable metabolite pair should be identified from the available data. We propose the following taxonomy:\\

\begin{enumerate}

	\item An ``optimally'' fit sample is one which falls inside a tolerance interval containing 80\% of the fresh-sample ratios with 95\% confidence (i.e., .80-content, .95-coverage TI).
	
	\item A ``desirably'' fit sample is one which falls inside a tolerance interval containing 95\% of the fresh-sample ratios with 95\% confidence (i.e., .95-content, .95-coverage TI) and which is not ``optimally'' fit.
	
	\item A ``minimally'' fit sample is one which falls inside a tolerance interval containing 99\% of the fresh-sample ratios with 95\% confidence (i.e., .99-content, .95-coverage TI) and which is not ``desirably'' fit.\\
	
\end{enumerate}

Tolerance intervals may be computed for data arising from approximately normal\cite{nist} or non-normal distributions\cite{Young2014}. In the case of approximately normal distributions, the two-sided $p$-content, $\gamma$-coverage tolerance interval is
\[
[\bar{Y}- s\cdot k,  \bar{Y} + s\cdot k],
\]
where $\bar{Y}$ is the sample mean, $s$ is the sample standard deviation, and $k$ is computed as:
\begin{equation} \label{tolerance1}
\displaystyle k = \sqrt{\dfrac{\nu\left( 1+\frac{1}{N} \right) z^2_{(1-p)/2}}{\chi^2_{1-\gamma, \nu}}},
\end{equation}

where $\chi^2_{1-\gamma, \nu}$ is the critical value of the chi-square distribution with degrees of freedom $\nu$ (usually $\nu=N-1$) that is exceeded with probability $\gamma$, and $z_{(1-p)/2}$ critical value of the normal distribution associated with cummulative probability (1-p)/2.\\

Changing-to-stable metabolite ratios offer a straightforward method for identifying fit and unfit samples. However, this method is not comprehensive: it may not identify all structure present in the available data for differentiating fresh and degraded samples. We therefore propose that changing-to-stable metabolite ratios may be used to train a probabilistic model which is capable of capturing additional structure in the data. We propose to utilize a Dirichlet process mixture model for this purpose (see Section \ref{dirichlet}) and also consider the use of k-Nearest-Neighbor and kernel regression methods (see Section \ref{kNN}).

\subsection{Modeling Methods} \label{models}

We discuss two main approach types for modeling metabolite sample fitness. The first approach type includes k-nearest neighbors (kNN) and kernel regression, two conceptually simple approaches not widely utilized for classification in the metabolomics literature. They are an intuitive method for classifying fit- and unfit samples, since fit samples should be more similar to other fit samples than unfit samples. However, kNN and kernel regression do not identify important relationships between significant features and are highly suceptible to misclassification due to contributions from insignificant features. Hence we also present a second approach.\\

The second approach type is that of Dirichlet process mixture models (DPMM), a non-parametric Bayesian approach\cite{Muller2004}. Despite the ``non-parametric'' descriptor, a key advantage of such models is not their absence of parameters but their flexible parametric form: the parametric form of the probability distributions is inferred along with the parameter values during model training. Characterization of metabolite sample fitness according to DPMM is highly intuitive because we regard sample feature measurements as arising from a combination of biological and chemical processes, including freshness (ideal sample collection and measurement) and degradation mechanisms including chemical (e.g. thermodynamic) and biological (e.g. bacterial) over varying lengths of time. Sample fitness may then be assessed according to the estimated probability of each process having generated a given sample.

\subsubsection{k-Nearest Neighbors and Kernel Regression} \label{kNN}


The k-Nearest Neighbor (kNN) classifier\cite{Bhatia2010, Thirumuruganathan2010} is one of the simplest classifiers conceptually and has only one parameter, $k$. Given training samples and a new test sample, it classifies the new sample in the majority category of its $k$ nearest training-sample neighbors. The nearness is computed according to a metric. Some metrics utilized for nearest neighbor classification include\cite{Bishop2007}:\\

\begin{enumerate}

	\item $L_2$ norm: normal Euclidean distance. The distance between two points is the square root of the sum of squared differences between the features.
	\item $L_1$ norm: The distance between two points is the sum of the absolute values of differences between the features.
	\item $L_{\infty}$ norm: The distance between two points is the absolute value of the greatest difference between any two features.\\
	
\end{enumerate}

The optimal $k$ may be determined using cross-validation. An advantage to using kNN for modeling fit and unfit samples is that it makes no parametric assumptions about the probability distributions of the sample groups or about the way in which those groups separate (in contrast to, for example, a hyperplane-separation method). However, since all features are treated equally in computing the distance between two points, the method is prone to mis-classification without exceptionally careful feature selection. K-nearest-neighbor classifiers have been discussed in the metabolomics literature but are rarely utilized for classification, in part due to their relatively large computational expense\cite{CuperlovicCulf2012}.\\

An extension of kNN is \emph{kernel regression}\cite{Bishop2007, Watson1964, Nadaraya1964}. kNN assumes that when classifying a new sample, all training samples should receive equal consideration and the number of neighbors (k) should be the same for classifying any new sample. Kernel regression relaxes those assumptions by defining a kernel which gives different weight to each training sample in the classification. Generally speaking, a \emph{kernel} is a similarity function which maps pairs of data points to a number; more similar (``nearer'') data points should be mapped to larger numbers. One widely-used kernel is a Gaussian kernel, which gives stronger weight to nearer training samples and less weight to farther-away training samples according to an exponential fall-off. For Gaussian kernel regression, the prediction $\hat{y}$ for a new sample $\mathbf{x}$ is
\begin{equation} \label{kernreg1}
\displaystyle \hat{y}(\mathbf{x}) = \argmax_{y \in Y} \sum_{i=1}^N I( f(\mathbf{x}_i) = y) K(\mathbf{x}_i, \mathbf{x}),
\end{equation}

where $y$ is a class identifier, $Y$ is the set of classes, $N$ is the number of training samples, $f$ maps a training sample to its class identifier, $I$ is the indicator function (I=1 if the argument is true, I=0 otherwise), and $K$ is the kernel function, in this case the Gaussian kernel\cite{Bishop2007}:
\begin{equation} \label{kernreg2}
\displaystyle K(\mathbf{x}_i, \mathbf{x}) = \dfrac{1}{\sqrt{(2\pi)^k\det{(\Sigma)}}}\exp{\left[-\frac{1}{2}(\mathbf{x_i}-\mathbf{x})^T\Sigma^{-1}(\mathbf{x_i}-\mathbf{x})\right]},
\end{equation}

where $k$ is the vector dimension of $\mathbf{x}$ (i.e. the number of features) and $\Sigma$ is a k x k covariance matrix. $\Sigma$ is a free parameter which may be determined by cross-validation or computed from the training data, e.g.
\begin{equation} \label{sigma}
\displaystyle \Sigma = \frac{1}{N-1}\sum_{i=1}^N (\mathbf{x}_i - \bar{\mathbf{x_i}})(\mathbf{x}_i - \bar{\mathbf{x_i}})^T,
\end{equation}

where $\bar{\mathbf{x_i}}$ is the average of the training samples.\\

Many kernels have their own parameters which must be determined by cross-validation; for example, the Gaussian kernel has a ``spread'' or ``standard deviation'' parameter $\sigma$ governing how fast the exponential falls off. As the number of parameters increases, so does the risk for over-fitting the prediction model.\\

\subsubsection{Dirichlet Process Mixture Model} \label{dirichlet}


We can conceptualize the sample fitness assessment problem as follows: we assume that samples are generated by a combination of distinct \emph{processes} (i.e. probability distributions) and consider those processes to be hidden variables in a mixture model. For example, we might consider a human urine sample metabolite levels to result from processes such as freshness (ideal fresh urine sample collection), bacterial contamination (due to exposure to above-freezing storage temperatures) over varying lengths of time, and chemical interactions within the urine (e.g. breakdown of thermodynamically unstable compounds or slow reactions) over varying lengths of time. Our proposed experimental procedure includes collection and measurement of biofluid metabolite levels after variable-time exposure to a range of storage temperatures. We propose to model the resulting data using a Dirichlet process mixture model (DPMM). To assess the fitness of a new sample, the DPMM will estimate the maximum posterior probability that the sample arose from each process. In this way we obtain a quantitative estimate of the degree to which a sample is generated from, e.g., freshness vs. various degradation processes. To make a sample fitness determination, we consider the following possibilities:\\

\begin{enumerate}

	\item Based on the principle of sample fitness (see Section \ref{principle}), the training samples may be used to establish a threshold for the probability of ``fresh'' process generation which constitutes a fit sample. Hence the final classifier predicts that the sample is fit or unfit based on the probability that the sample arose from the ``fresh'' process.\\
	
	\item However, more than one process may be associated with sample ``freshness''. Therefore it may be desirable to use the process probabilities computed by the DPMM as inputs to a final classifier for sample fitness. For this procedure, we will explore the use of classifiers such as kNN, O-PLS-DA\cite{Bylesjo2007}, and hyperplane methods such as support vector machine\cite{Srivastava2005} or soft independent modeling of class analogy\cite{Wold1977} (SIMCA). Hence the final classifier assess sample fitness using the full vector of process probabilities computed by the DPMM.\\
	
\end{enumerate}

The advantage of using a DPMM lies in its parametric flexibility: the priors for each process are defined over a space of functions, and the appropriate parametric representation of each process is inferred \emph{together with} the parameter values. In contrast, for more strongly-parametrized approaches such as a Gaussian Mixture Model, the parametric representation of each process is assumed and only the parameter values are computed in model training.\\

\bibliography{references}{}

\begin{thebibliography}{10}

\bibitem{Keun2011}
Keun H.C. and Athersuch T.J.
\newblock {Nuclear magnetic resonance (NMR)-based metabolomics}.
\newblock {\em Methods Mol Biol.}, 2011.

\bibitem{Beckonert2007}
Beckonert~O. et~al.
\newblock {Metabolic profiling, metabolomic and metabonomic procedures for NMR
  spectroscopy of urine, plasma, serum and tissue extracts}.
\newblock {\em Nature protocols 2 pp. 2692-703}, 2007.

\bibitem{Zhou2012}
Zhou~B. et~al.
\newblock {LC-MS-based metabolomics.}
\newblock {\em Mol Biosyst.}, 2012.

\bibitem{Want2013}
Want~E et~al.
\newblock {Global metabolic profiling of animal and human tissues via UPLC-MS.}
\newblock {\em Nature Protocols}, 2013.

\bibitem{Benton2012}
Benton~H.P. et~al.
\newblock {Intra- and Interlaboratory Reproducibility of Ultra Performance
  Liquid Chromatography–Time-of-Flight Mass Spectrometry for Urinary Metabolic
  Profiling}.
\newblock {\em Anal. Chem. 84 (5)}, 2012.

\bibitem{Razali2011}
Razali N.H. and Yap B.
\newblock {Power comparisons of Shapiro-Wilk, Kolmogorov-Smirnov, Lilliefors
  and Anderson-Darling tests}.
\newblock {\em Journal of Statistical Modeling and Analytics}, 2011.

\bibitem{Weljie2006}
Weljie~A.M. et~al.
\newblock {Targeted profiling: quantitative analysis of 1H NMR metabolomics
  data.}
\newblock {\em Anal Chem.}, 2006.

\bibitem{Athersuch2013}
Athersuch~T.J. et~al.
\newblock {Evaluation of 1H NMR metabolic profiling using biofluid mixture
  design.}
\newblock {\em Anal Chem.}, 2013.

\bibitem{Weljie2008}
Weljie~A.M. et~al.
\newblock {Evaluating low-intensity unknown signals in quantitative proton NMR
  mixture analysis.}
\newblock {\em Anal. Chem.}, 2008.

\bibitem{Tredwell2011}
Tredwell~G.D. et~al.
\newblock {Between-Person Comparison of Metabolite Fitting for NMR-Based
  Quantitative Metabolomics}.
\newblock {\em Anal. Chem.}, 2011.

\bibitem{Keun2002}
Keun~H.C. et~al.
\newblock {Analytical reproducibility in (1)H NMR-based metabonomic
  urinalysis}.
\newblock {\em Chem. Res. Toxicol.}, 2002.

\bibitem{Ravanbakhsh2014}
Ravanbakhsh~S. et~al.
\newblock {Accurate, fully-automated NMR spectral profiling for metabolomics}.
\newblock {\em arXiv}, 2014.

\bibitem{Lacy2014}
Lacy~P. et~al.
\newblock {Signal Intensities Derived from Different NMR Probes and Parameters
  Contribute to Variations in Quantification of Metabolites.}
\newblock {\em PLoS One}, 2014.

\bibitem{Zheng2013}
Zheng~H. et~al.
\newblock {Time-Saving Design of Experiment Protocol for Optimization of LC-MS
  Data Processing in Metabolomic Approaches.}
\newblock {\em Anal. Chem.}, 2013.

\bibitem{Xiao2012}
Xiao~J.F. et~al.
\newblock {Metabolite identification and quantitation in LC-MS/MS-based
  metabolomics}.
\newblock {\em Trends Analyt Chem.}, 2012.

\bibitem{Oberacher2009}
Oberacher~H. et~al.
\newblock {On the inter-instrument and inter-laboratory transferability of a
  tandem mass spectral reference library: 1. Results of an Austrian multicenter
  study}.
\newblock {\em J Mass Spectrom.}, 2009.

\bibitem{Dona2014}
Dona~A.C. et~al.
\newblock {Precision high-throughput proton NMR spectroscopy of human urine,
  serum, and plasma for large-scale metabolic phenotyping}.
\newblock {\em Anal Chem.}, 2014.

\bibitem{Emwas2014}
Emwas~A. et~al.
\newblock {Standardizing the experimental conditions for using urine in
  NMR-based metabolomic studies with a particular focus on diagnostic studies:
  a review}.
\newblock {\em Metabolomics}, 2014.

\bibitem{Kamlage2014}
Kamlage~B. et~al.
\newblock {Quality Markers Addressing Preanalytical Variations of Blood and
  Plasma Processing Identified by Broad and Targeted Metabolite Profiling}.
\newblock {\em Clinical Chemistry}, 2014.

\bibitem{Yin2013}
Yin~P. et~al.
\newblock {Preanalytical Aspects and Sample Quality Assessment in Metabolomics
  Studies of Human Blood.}
\newblock {\em Clinical Chemistry}, 2013.

\bibitem{Korman2012}
Korman~A. et~al.
\newblock {Statistical methods in metabolomics}.
\newblock {\em Methods Mol Biol.}, 2012.

\bibitem{Bruce2009}
Bruce~S.J. et~al.
\newblock {Investigation of Human Blood Plasma Sample Preparation for
  Performing Metabolomics Using Ultrahigh Performance Liquid
  Chromatography/Mass Spectrometry}.
\newblock {\em Anal. Chem.}, 2009.

\bibitem{Saude2007}
Saude E.J. and Sykes B.D.
\newblock {Urine stability for metabolomic studies: effects of preparation and
  storage.}
\newblock {\em Metabolomics}, 2007.

\bibitem{Elliot2007}
Elliot P. and Peakman T.C.
\newblock {The UK Biobank sample handling and storage protocol for the
  collection, processing and archiving of human blood and urine.}
\newblock {\em Int. J. Epidemiol.}, 2008.

\bibitem{Dunn2008}
Dunn~W.B. et~al.
\newblock {A GC-TOF-MS study of the stability of serum and urine metabolomes
  during the UK Biobank sample collection and preparation protocols}.
\newblock {\em Int J Epidemiol.}, 2008.

\bibitem{Hebels2013}
Hebels~D.G.A.J. et~al.
\newblock {Performance in Omics Analyses of Blood Samples in Long-Term Storage:
  Opportunities for the Exploitation of Existing Biobanks in Environmental
  Health Research}.
\newblock {\em Environ Health Perspect}, 2013.

\bibitem{Giraudeau2014}
Giraudeau P., Tea I., Remaud G., and Akoka S.
\newblock {Reference and normalization methods: Essential tools for the
  intercomparison of NMR spectra.}
\newblock {\em Journal of Pharmaceutical and Biomedical Analysis}, 2014.

\bibitem{Ejigu2013}
Ejigu~B.A. et~al.
\newblock {Evaluation of Normalization Methods to Pave the Way Towards
  Large-Scale LC-MS-Based Metabolomics Profiling Experiments}.
\newblock {\em OMICS}, 2013.

\bibitem{Wang2013}
Wang B., A.~Goodpaster, and Kennedy M.
\newblock {Coefficient of variation, signal-to-noise ratio, and effects of
  normalization in validation of biomarkers from NMR-based metabonomics
  studies}.
\newblock {\em Chemometrics and Intelligent Laboratory Systems}, 2013.

\bibitem{Slaff2014}
Slaff B., Sengupta A., and Weljie A.M.
\newblock {(In Press.) NMR Spectroscopy of Urine.}
\newblock In Keun H.C., editor, {\em NMR-based Metabolomics}. 2014.

\bibitem{Dieterle2006}
Dieterle et~al.
\newblock {Probabilistic quotient normalization as robust method to account for
  dilution of complex biological mixtures. Application in 1H NMR metabonomics.}
\newblock {\em Anal. Chem.}, 2006.

\bibitem{Veselkov2011}
Veselkov~KA et~al.
\newblock {Optimized preprocessing of ultra-performance liquid
  chromatography/mass spectrometry urinary metabolic profiles for improved
  information recovery}.
\newblock {\em Anal Chem.}, 2011.

\bibitem{Maesschalck2000}
De~Maesschalck~R. et~al.
\newblock {The Mahalanobis distance}.
\newblock {\em Chemometrics and Intelligent Laboratory Systems}, 2005.

\bibitem{Geitz2007}
Geitz.
\newblock {Vector Geometry for Computer Graphics}.
\newblock {\em
  https://www.cs.oberlin.edu/~bob/cs357.08/VectorGeometry/VectorGeometry.pdf}.

\bibitem{vandenBerg2006}
van den Berg R. A.~et al.
\newblock {Centering, scaling, and transformations: improving the biological
  information content of metabolomics data.}
\newblock {\em BMC Genomics}, 2006.

\bibitem{Gromsky2014}
Gromsky~P.S. et~al.
\newblock {The influence of scaling metabolomics data on model classification
  accuracy}.
\newblock {\em Metabolomics}, 2015.

\bibitem{Keun2003}
Keun~H.C. et~al.
\newblock {Improved analysis of multivariate data by variable stability
  scaling: application to NMR-based metabolic profiling}.
\newblock {\em Analytica Chimica Acta}, 2003.

\bibitem{Riter2005}
Riter~L. et~al.
\newblock {Statistical design of experiments as a tool in mass spectrometry}.
\newblock {\em J Mass Spectrom}, 2005.

\bibitem{Eliasson2012}
Eliasson~M et~al.
\newblock {Strategy for Optimizing LC-MS Data Processing in Metabolomics: A
  Design of Experiments Approach}.
\newblock {\em Anal. Chem.}, 2012.

\bibitem{KorkmazMVN}
Korkmaz, Goksuluk, and Zararsiz.
\newblock {MVN: An R Package for Assessing Multivariate Normality}.
\newblock {\em http://cran.r-project.org/web/packages/MVN/vignettes/MVN.pdf}.

\bibitem{LillieforsH}
H.~Lilliefors.
\newblock {On the Kolmogorov Smirnov test for normality with mean and variance
  unknown}.
\newblock {\em Journal of the American Statistical Association}, 1967.

\bibitem{Moseley2013}
Moseley H.N.B.
\newblock {Error Analysis and Propagation in Metabolomics Data Analysis}.
\newblock {\em Comput Struct Biotechnol J.}, 2013.

\bibitem{Gross2015}
Gross J. and Ligges U.
\newblock {nortest: Five omnibus tests for testing the composite hypothesis of
  normality.}
\newblock {\em http://cran.r-project.org/web/packages/nortest/nortest.pdf}.

\bibitem{Saeys2007}
Saeys Y., Inza I., , and Larranaga P.
\newblock {A review of feature selection techniques in bioinformatics.}
\newblock {\em Bioinformatics}, 2007.

\bibitem{delosCampos2013}
de~los Campos G.~et al.
\newblock {Whole-Genome Regression and Prediction Methods Applied to Plant and
  Animal Breeding}.
\newblock {\em Genetics}, 2013.

\bibitem{Bartel2013}
Bartel J., Krumsiek J., and Theis F.J.
\newblock {Statistical methods for the analysis of high-throughput metabolomics
  data}.
\newblock {\em Comput Struct Biotechnol J.}, 2013.

\bibitem{Wold2001}
Wold S.~Sjostrom M. and Eriksson L.
\newblock {PLS-regression: a basic tool of chemometrics}.
\newblock {\em Chemometrics and Intelligent Laboratory Systems}, 2001.

\bibitem{Hardoon2004}
Hardoon D.R., Szedmak S., and Shawe-Taylor J.
\newblock {Canonical correlation analysis; An overview with application to
  learning methods.}
\newblock {\em Neural Computation}, 2004.

\bibitem{Trygg2002opls}
Trygg J. and Wold S.
\newblock {Orthogonal Projections to Latent Structures (O-PLS).}
\newblock {\em J. Chemometrics}, 2002.

\bibitem{Bylesjo2007}
Bylesjö~M. et~al.
\newblock {OPLS discriminant analysis: combining the strengths of PLS-DA and
  SIMCA classification}.
\newblock {\em Journal of Chemometrics}, 2006.

\bibitem{Trygg2002o2}
Trygg J.
\newblock {O2-PLS for qualitative and quantitative analysis in multivariate
  calibration.}
\newblock {\em Journal of Chemometrics}, 2002.

\bibitem{Madsen2010}
Madsen R., Lundstedt T., and Trygg J.
\newblock {Chemometrics in metabolomics-A review in human disease diagnosis.}
\newblock {\em Analytica Chimica Acta}, 2010.

\bibitem{Varmuza2009}
Varmuza K. and Filzmoser P.
\newblock {\em {Introduction to Multivariate Statistical Analysis in
  Chemometrics}}.
\newblock 2009.

\bibitem{Tibshirani2002}
Tibshirani~R. et~al.
\newblock {Diagnosis of multiple cancer types by shrunken centroids of gene
  expression.}
\newblock {\em Proceedings of the National Academy of Sciences of the United
  States of America}, 2002.

\bibitem{Sha2010}
Sha~W. et~al.
\newblock {Metabolomic profiling can predict which humans will develop liver
  dysfunction when deprived of dietary choline.}
\newblock {\em FASEB}, 2010.

\bibitem{Chen2014}
Chen~C. et~al.
\newblock {Shrunken centroids regularized discriminant analysis as a promising
  strategy for metabolomics data exploration}.
\newblock {\em Journal of Chemometrics}, 2015.

\bibitem{Krier2007}
Krier~C. et~al.
\newblock {Feature clustering and mutual information for the selection of
  variables in spectral data.}
\newblock {\em European Symposium on Artificial Neural Networks}, 2007.

\bibitem{Magendiran2014}
Magendiran N. and Jayaranjani J.
\newblock {An Efficient Fast Clustering-Based Feature Subset Selection
  Algorithm for High Dimensional Data.}
\newblock {\em International Journal of Innovative Research in Science,
  Engineering, and Technology}, 2014.

\bibitem{Srivastava2005}
Srivastava D.K. and Bhambhu L.
\newblock {Data Classification Using Support Vector Machine.}
\newblock {\em Journal of Theoretical and Applied Information Technology},
  2005.

\bibitem{Bhatia2010}
Bhatia N. and Vandana A.
\newblock {Survey of Nearest Neighbor Techniques}.
\newblock {\em International Journal of Computer Science and Information
  Security}, 2010.

\bibitem{Boulesteix2012}
Boulesteix~A. et~al.
\newblock {Overview of Random Forest Methodology and Practical Guidance with
  Emphasis on Computational Biology and Bioinformatics}.
\newblock {\em Wiley Interdisciplinary Reviews: Data Mining and Knowledge
  Discovery}, 2012.

\bibitem{Kim2008}
Kim~S.B. et~al.
\newblock {Controlling the False Discovery Rate for Feature Selection in
  High-resolution NMR Spectra}.
\newblock {\em Stat Anal Data Min.}, 2008.

\bibitem{Fraser1997}
Fraser~C.G. et~al.
\newblock {Proposal for setting generally applicable quality goals solely based
  on biology.}
\newblock {\em Ann Clin Biochem}, 1997.

\bibitem{Plebani2015}
Plebani~M. et~al.
\newblock {Performance criteria and quality indicators for the pre-analytical
  phase}.
\newblock {\em Clin Chem Lab Med}, 2015.

\bibitem{Wald1946}
Wald A. and Wolfowitz J.
\newblock {Tolerance limits for a normal distribution.}
\newblock {\em The Annals of Mathematical Statistics.}, 1946.

\bibitem{Brown2009}
Brown S.D., Ferré R.T.I., and Walczak B.
\newblock {\em {Comprehensive Chemometrics: Statistics, experimental design,
  optimization.}}
\newblock 2009.

\bibitem{nist}
J.~Prins.
\newblock {7.2.6. What intervals contain a fixed percentage of the population
  values?}
\newblock In Croarkin C. and Tobias P., editors, {\em NIST/SEMATECH e-Handbook
  of Statistical Methods}. 2012.

\bibitem{Young2014}
Young D.S. and Mathew T.
\newblock {Improved nonparametric tolerance intervals based on interpolated and
  extrapolated order statistics.}
\newblock {\em Journal of Nonparametric Statistics}, 2014.

\bibitem{Muller2004}
Muller P. and Quintana~F. A.
\newblock {Nonparametric Bayesian Data Analysis}.
\newblock {\em Statistical Science}, 2004.

\bibitem{Thirumuruganathan2010}
Thirumuruganathan S.
\newblock {A Detailed Introduction to K-Nearest Neighbor (KNN) Algorithm.}
\newblock {\em
  https://saravananthirumuruganathan.wordpress.com/2010/05/17/a-detailed-introduction-to-k-nearest-neighbor-knn-algorithm/}.

\bibitem{Bishop2007}
Bishop C.
\newblock {\em {Pattern Recognition and Machine Learning}}.
\newblock Springer, 2007.

\bibitem{CuperlovicCulf2012}
Cuperlovic-Culf M.
\newblock {\em {NMR Metabolomics in Cancer Research}}.
\newblock Woodhead Publishing, 2012.

\bibitem{Watson1964}
Watson G.S.
\newblock {Smooth Regression Analysis.}
\newblock {\em Sankhya: The Indian Journal of Statistics}, 1964.

\bibitem{Nadaraya1964}
Nadaraya E.A.
\newblock {On Estimating Regression}.
\newblock {\em Theory Probab. Appl.}, 1964.

\bibitem{Wold1977}
Wold S. and Sjostrom M.
\newblock {SIMCA: A method for analyzing chemical data in terms of similarity
  and analogy}.
\newblock In Kowalski B.R., editor, {\em Chemometrics Theory and Application}.
  1977.

\end{thebibliography}
\bibliographystyle{unsrt}

\end{document}